\documentclass[useAMS,usenatbib]{mn2e}
\usepackage{graphicx}
\usepackage{amsmath}
\title[Solar System chemical abundances]{Solar System chemical abundances corrected for systematics}
\author[G.\ Gonzalez]{Guillermo Gonzalez$^{1}$\\
$^{1}$Department of Physics and Astronomy, Ball State University, Muncie, IN 47306 USA\\
}

\begin{document}

\date{Accepted ??. Received ??; in original form ??}

\pagerange{\pageref{firstpage}--\pageref{lastpage}} \pubyear{??}

\maketitle

\label{firstpage}

\begin{abstract}
The relative chemical abundances between CI meteorites and the solar photosphere exhibit a significant trend with condensation temperature. A trend with condensation temperature is also seen when the solar photospheric abundances are compared to those of nearby solar twins. We use both these trends to determine the alteration of the elemental abundances of the meteorties and the photosphere by fractionation and calculate a new set of primordial Solar System abundances.
\end{abstract}

\section{Introduction}

The Solar System chemical abundances are derived primarily from two sources: CI chondrite meteorites and the solar photosphere \citep{lod09}. \citet{gg10} showed that their relative abundances display a trend with condensation temperature (T$_{\rm c}$), in the sense that the high T$_{\rm c}$ elements in the solar photosphere are more abundant than they are in meteorites. However, it is not yet been demonstrated whether the source of this trend is the abundance pattern of the photosphere, meteorites or both. 

Recently, an extensive suite of solar photospheric chemical abundances have been redetermined by \citet{grev14} and \citet{scot14a,scot14b} from a careful consideration of all ingredients in the abundance analysis (3D solar atmosphere model, NLTE line formation, selection of lines, atomic/molecular data, etc). We will use these new data to re-examine the meteorite-photosphere relative abundance-T$_{\rm c}$ trend and compare it to the abundance-T$_{\rm c}$ trend between the Sun and nearby Sun-like stars that have been reported in the literature. In particular, we will quantify the amount of fractionation that the CI chondrite meteorites have experienced relative to the primordial Solar System abundances. We will then use these results to present a new set of meteoritic and photospheric abundances, each set corrected for fractionation effects.

\section{Comparing Photospheric and meteoritic abundances}

\subsection{Relative meteoritic-photospheric abundance trend with T$_{\rm c}$}

\citet{gg97} first noted a trend between the difference in the Solar System photospheric and meteoritic abundances and T$_{\rm c}$. In \citet{gg06} we revisited this topic using the abundance data of \citet{lod03} and \citet{asp05} and confirmed the trend. We repeated the analysis in \citet{gg10} with phototospheric data from \citet{asp09} and meteoritic data from \citet{lod09}. For the present study we make use of the same meteoritic data as in \citet{gg10}, adjusted to match the Si abundance in \citet{scot14a}) and the photospheric abundances from \citet{grev14} and \citet{scot14a,scot14b}. 

We show in Figure 1 the difference between the solar photospheric and meteoritic abundances ($\Delta_{\rm pm} = \log $(X/H)$_{\odot \rm p} - \log$ (X/H)$_{\odot \rm m}$) plotted against T$_{\rm c}$. The figure includes the 50 elements with quoted individual abundance uncertainties $\le 0.10$ dex. The errors bars for each element are calculated from the quadrature sum of the individual photospheric and meteoritic uncertainties. The slope from a weighted linear least-squares solution is $(7.4 \pm 2.5) \times10^{\rm -5}$ dex K$^{\rm -1}$, and the adjusted $R^{\rm 2}$ value is 0.12.\footnote{The goodness of fit $R^{\rm 2}$ value, also called the {\it coefficient of determination}, is the square of the Pearson correlation coefficient in the case of linear least-squares. The adjusted $R^{\rm 2}$ value takes into account the number of predictors in the model and is always less than the $R^{\rm 2}$ value.} This result is nearly identical to that obtained by \citet{gg10}.

Although there is considerable scatter evident in the figure, it is important to note that the data points with the largest deviations from the trend line tend to have larger uncertainties, and thus receive less weight in the fit. To test the significance of this result, we produced 10 000 random reorderings of the T$_{\rm c}$ values associated with each $\Delta_{\rm pm}$ value and calculated the adjusted $R^{\rm 2}$ for each case. Only 0.37\% of the simulated datasets produced larger adjusted $R^{\rm 2}$ values than the actual dataset.

\begin{figure}
  \includegraphics[width=3.3in]{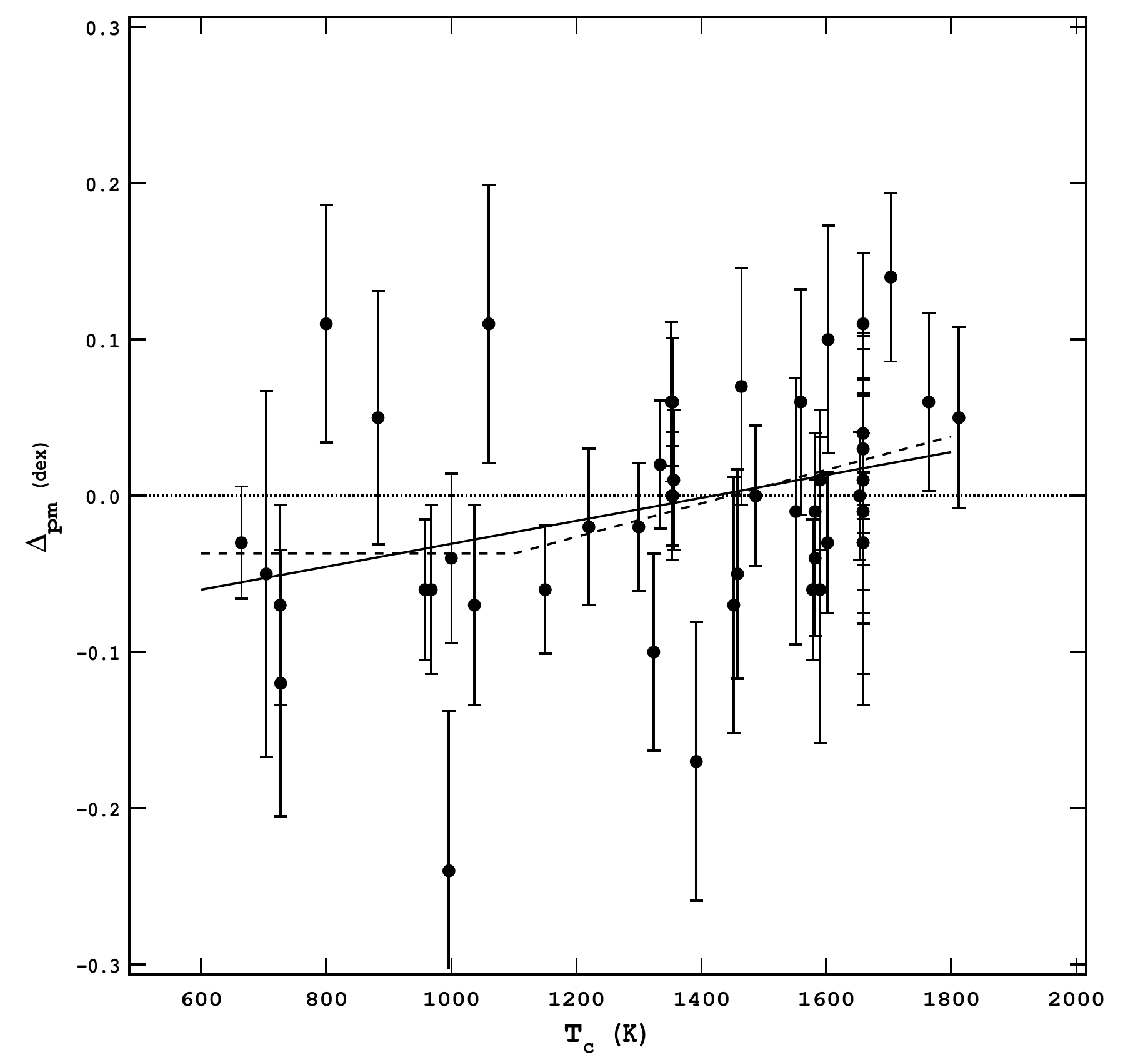}
 \caption{Difference between the solar photospheric and meteoritic elemental abundances ($\Delta_{\rm pm}$) is plotted against condensation temperature (T$_{\rm c}$). The error bars include the errors from both the solar and meteoritic abundances. The solid line is a weighted least-squares linear fit. The dashed line is the third model described in the text, with a break at 1100 K.}
\end{figure}

It is also helpful to compare the linear fit shown in Figure 1 to one determined by fitting a dataset with the T$_{\rm c}$ values replaced with the atomic number of each element. We have done so and show the result in Figure 2. The y-intercept and slope values determined from a weighted linear least-squares fit are $-0.009 \pm 0.016$ dex and $(1.9 \pm 3.7) \times10^{\rm -4}$ dex K$^{\rm -1}$, respectively; the adjusted $R^{\rm 2}$ value is $-0.035$.

\begin{figure}
  \includegraphics[width=3.3in]{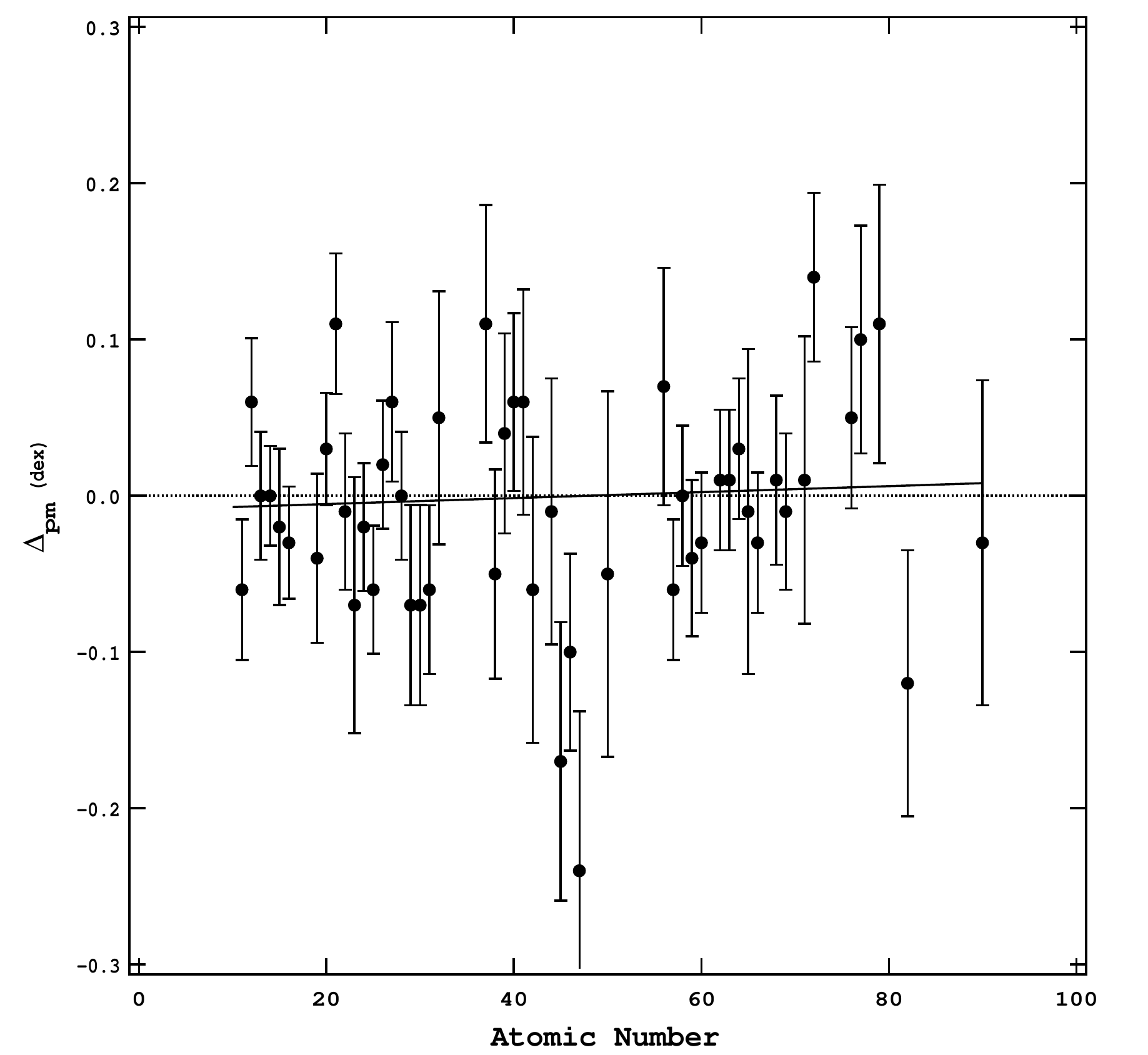}
 \caption{$\Delta_{\rm pm}$ versus atomic number for the same elements in Figure 1. The solid line is a linear least-squares fit.}
\end{figure}

Given the scatter in the data evident in Figure 1, it is not clear whether a simple linear trend over the full range of T$_{\rm c}$ values is the best model to account for the variance. We also fit a second model, one which assumes a simple step at $T_{\rm c} = 1100$ K, with constant values on either side. This model gives an adjusted $R^{\rm 2}$ value of 0.06, worse than the simple linear model.

Finally, in a third model, we assume a constant value of $\Delta_{\rm pm}$ below T$_{\rm c} = 1100$, and a linear trend above 1100 K. We find that the constant is $-0.037 \pm 0.013$ dex and the slope above 1100 K is $(10.7 \pm 3.3) \times10^{\rm -5}$ dex K$^{\rm -1}$; the adjusted $R^{\rm 2}$ value is 0.14. Thus, this model has a slightly better goodness of fit than the other two models. Unfortunately, we are not able to set tight limits on the location of the break; plausible values range from about 1000 to 1400 K. To test the sensitivity of the resultant solution to the location of the break point, we also performed a fit with a break at T$_{\rm c} = 1250$ (which we employ below). The constant and slope values for this case are $-0.029 \pm 0.012$ dex and $(12.7 \pm 4.2) \times10^{\rm -5}$ dex K$^{\rm -1}$, respectively; the adjusted $R^{\rm 2}$ value is 0.12.

\section{Discussion}

We showed in the previous section that the pattern of $\Delta_{\rm pm}$ with T$_{\rm c}$ is best explained with a model having a constant value below T$_{\rm c} \sim 1100$ K, and a linear trend above it. While other models can also fit the data, we are not justified in using anything more complex, given the level of scatter.

This pattern is similar to that reported for nearby solar twins \citep{mel09,mel12}. Figure 3 of \citet{mel09}, a plot of solar abundances minus the average of solar twins abundances, shows a break at T$_{\rm c} = 1200$ K. They fit a linear trend on each side of the break; the magnitude of the slope for the high-T$_{\rm c}$ side is $9.5 \times10^{\rm -5}$ dex K$^{\rm -1}$. This value is consistent with our fit. The data on the cool side of the break could just as well be fit with a constant value.

In their study of the best-matching solar twin discovered to date, HIP 56948, \citet{mel12} found a similar pattern. However, they found a more shallow slope, $2.1 \times10^{\rm -5}$ dex K$^{\rm -1}$, and a cooler break point, $T_{\rm c} \sim 1000$ K.

These comparisons show that the same general pattern in the relative abundances between the photosphere and meteorites is also present in the relative abundances between the Sun and nearby solar twins. One is tempted to ascribe the source of this pattern to the Sun's photosphere. Indeed, \citet{mel09} proposed that this pattern is a fingerprint of early planet formation processes in the Solar System. Specifically, they suggested that as the terrestrial planets were forming, the gas in the inner Solar System was partially depleted of highly refractory elements. The depleted gas was eventually accreted onto the Sun, where it left its imprint on the abundance pattern of its photosphere. \citet{cham10} quantified this idea further, finding that the solar twins abundance pattern can be matched with the Sun's abundances if 4 Earth masses of Earth-like and carbonaceous chondrite-like material are added to the Sun's photosphere. This implies that this mix and quantity of material had been removed from the gas in the protoplanetary disk before it was accreted onto the Sun.

This seems like an attractive explanation for the Solar System relative abundance pattern reported in the present work, but the sign of the slope is wrong. The refractory element abundances are enhanced in the photosphere relative to the meteorites! One would expect to find the reverse of this pattern if the Melendez et al. hypothesis is correct. In order to salvage it, one would have to propose either that the meteorites have been depleted in refractory elements as well or that the meteorties are enhanced in volatiles. It is difficult to choose between these ideas without additional evidence for one or the other.

There is a third source of Solar System abundances, $s$-process systematics. \citet{grev14} have adopted theoretical $s$-process values for the Solar System abundances of Kr and Xe, which lack photospheric spectral lines. $s$-process systematics can also be useful in studies of the meteoritic isotopic abundance patterns. In particular, several isotopes are known to be produced only by the $s$-process. These are especially valuable, because they avoid having to know the contributions from the much less well-understood $r$- and $p$-processes. 

\citet{bis14} have published the results of detailed $s$-process calculations based on a Galactic chemical evolution model and detailed stellar physics to produce a best match to the Solar System isotopic abundance pattern (based on meteorite data). They noted a few anomalies relative to their best fit case. First, their model underestimates the $s$-only isotope abundances relative to the Solar System values between A = 90 and 130 by about 25\%.\footnote{It is important to note that this deficit of light $s$-process isotopes is relative to the abundance of the $s$-only isotope $^{150}$Sm, which they have chosen as a reference for reporting their abundances.} They also underestimate the abundance of the $s$-only isotope $^{204}$Pb by about 13\%. The neutron capture cross sections of the Pb isotopes are well known, but the cross section of $^{\rm 204}$Tl, which affects the $^{204}$Pb abundance through a branching, is not as well determined (only theoretical estimates exist). \citet{bis14} suggested a missing $s$-process component and neutron capture cross section errors to account for these anomalies. Some could be explained by fractionation of the volatile versus refractory elements in meteorites. 

Another approach is to forego the attempt to establish the primordial Solar System abundances using only Solar System data and, instead, bring in data external to the Solar System. This may be our best option, given the evidence that both the solar photospheric and meteoritic abundances are not accurate samples of the primordial abundances. An obvious choice for the external reference is the mean abundance pattern of the nearby solar twins determined by \citet{mel09}. We will not use the detailed element-to-element pattern as the reference but only the trend with T$_{\rm c}$. In this way, we can preserve the detailed Solar System abundance pattern while at the same time removing the trends with T$_{\rm c}$. Implicit in this approach is the assumption that the average abundance pattern of the solar twins has not been affected by fractionation processes.

We use the abundance trend with $T_{\rm c}$ from Figure 3 of \citet{mel09}, which is:
\begin{eqnarray*}
[X/Fe]_{\rm cpf} = \left\{ \begin{array}{lr}
-0.052 + 2.4 \times 10^{-5} T_{\rm c} &\mbox{ if $T_{\rm c} \le 1250$ K}\\
\\
-0.022 + 10^{-4} (T_{\rm c} - 1250) &\mbox{if $T_{\rm c} > 1250$ K}\\
       \end{array} \right.
\end{eqnarray*}
where [X/Fe]$_{\rm cpf}$ is the fitted log of the abundance ratio of element X to Fe in the solar twin stars minus the log of the same ratio in the Sun's photosphere ($\log$ (X/Fe)$_{\rm cp} - \log$ (X/Fe)$_{\odot \rm p}$); here, `p' stands for `photosphere', `c' stands for `comparison' and `f' stands for `fitted.' Using this equation, then, we can remove the abundance-T$_{\rm c}$ pattern from the solar photosphere by adding [X/Fe]$_{\rm cpf}$ to the solar photospheric abundances (as X/Fe) according to each element's value of T$_{\rm c}$. In doing this, we are implicitly equating the primordial (corrected) solar photospheric abundances to $\log$ (X/Fe)$_{\rm cp}$. In equation form,
\begin{eqnarray*}
\log (X/Fe)^{\star}_{\odot \rm p}=\log (X/Fe)_{\odot \rm p} + [X/Fe]_{\rm cpf}
\end{eqnarray*}
where the asterisk signifies the corrected abundance ratio. These abundance ratios can be converted into $\log$ (X/H) abundance values by adding 7.47 dex, which is the present solar photospheric Fe abundance value \citep{scot14b}. The next step is to correct for the relative depletion of Fe between the Sun and the solar twins. According to \citet{mel09} and \citet{cham10} the Sun's photosphere is depleted in Fe by about 0.04 dex relative to the volatile elements relative to the solar twins. If we assume the volatile elements in the Sun and solar twins have not been fractionated, then we can add 0.04 dex to the solar photospheric Fe abundance to correct for its depletion during planet formation. A metal-depleted solar envelope relative to the solar interior would also help to resolve the continuing discrepancy between the spectrocopically-determined photospheric metallicity and helioseismology \citep{bas13}. Finally, to determine the primordial abundances, another 0.03 dex needs to be added to all the abundance values to correct for the effects of diffusion in the Sun's atmosphere over the course of its history \citep{mow12}.

The meteoritic abundances can be corrected by a similar procedure. We begin by defining several quantities:
\begin{align*}
[X/Fe]_{\rm cm} &=\log (X/Fe)_{\rm cp} - \log (X/Fe)_{\odot \rm m}\\
\Delta_{\rm pm} &= \log (X/H)_{\odot \rm p} - \log (X/H)_{\odot \rm m}\\
\Delta_{\rm Fe} &= \log (Fe/H)_{\odot \rm p} - \log (Fe/H)_{\odot \rm m}\\
\Delta_{\rm pmFe} &= \log (X/Fe)_{\odot \rm p} - \log (X/Fe)_{\odot \rm m}=\Delta_{\rm pm} - \Delta_{\rm Fe}
\end{align*}
where the `m' subscript refers to the meteoritic value, and the quantity $\Delta_{\rm Fe}$ is equal to 0.02 dex. From these quantities we can solve for [X/Fe]$_{\rm cmf}$:
\begin{eqnarray*}
[X/Fe]_{\rm cmf}=[X/Fe]_{\rm cpf} + \Delta_{\rm pmf} - \Delta_{\rm Fe}
\end{eqnarray*}
In order to simplify the equation for [X/Fe]$_{\rm cmf}$, we set the $T_{\rm c}$ break value to be 1250 K for $\Delta_{\rm pmf}$. The new equation for $\Delta_{\rm pmf}$ is:
\begin{eqnarray*}
\Delta_{\rm pmf} = \left\{ \begin{array}{lr}
-0.029 &\mbox{ if $T_{\rm c} \le 1250$ K}\\
\\
-0.029~+ \\
1.27 \times 10^{-4} (T_{\rm c} - 1250) &\mbox{if $T_{\rm c} > 1250$ K}\\
       \end{array} \right.
\end{eqnarray*}
The resulting equation for [X/Fe]$_{\rm cmf}$ is:
\begin{eqnarray*}
[X/Fe]_{\rm cmf} = \left\{ \begin{array}{lr}
-0.101 + 2.4 \times 10^{-5} T_{\rm c} &\mbox{ if $T_{\rm c} \le 1250$ K}\\
\\
-0.071~+ \\
2.27 \times 10^{-4} (T_{\rm c} - 1250) &\mbox{if $T_{\rm c} > 1250$ K}\\
       \end{array} \right.
\end{eqnarray*}
Finally, we can calculate the corrected Solar System meteritic abundance ratios from:
\begin{eqnarray*}
\log (X/Fe)^{\star}_{\odot \rm m}=\log (X/Fe)_{\odot \rm m} + [X/Fe]_{\rm cmf}
\end{eqnarray*}
This quantity can be converted into $\log$ (X/H) by adding, 7.45, which is the present meteoritic Fe abundance. In addition, we added a constant offset (0.08 dex) to the meteoritic abundances so that the meteoritic Si abundance matches the primordial Si photospheric abundance, which was calculated as described above.

\begin{figure}
  \includegraphics[width=3.3in]{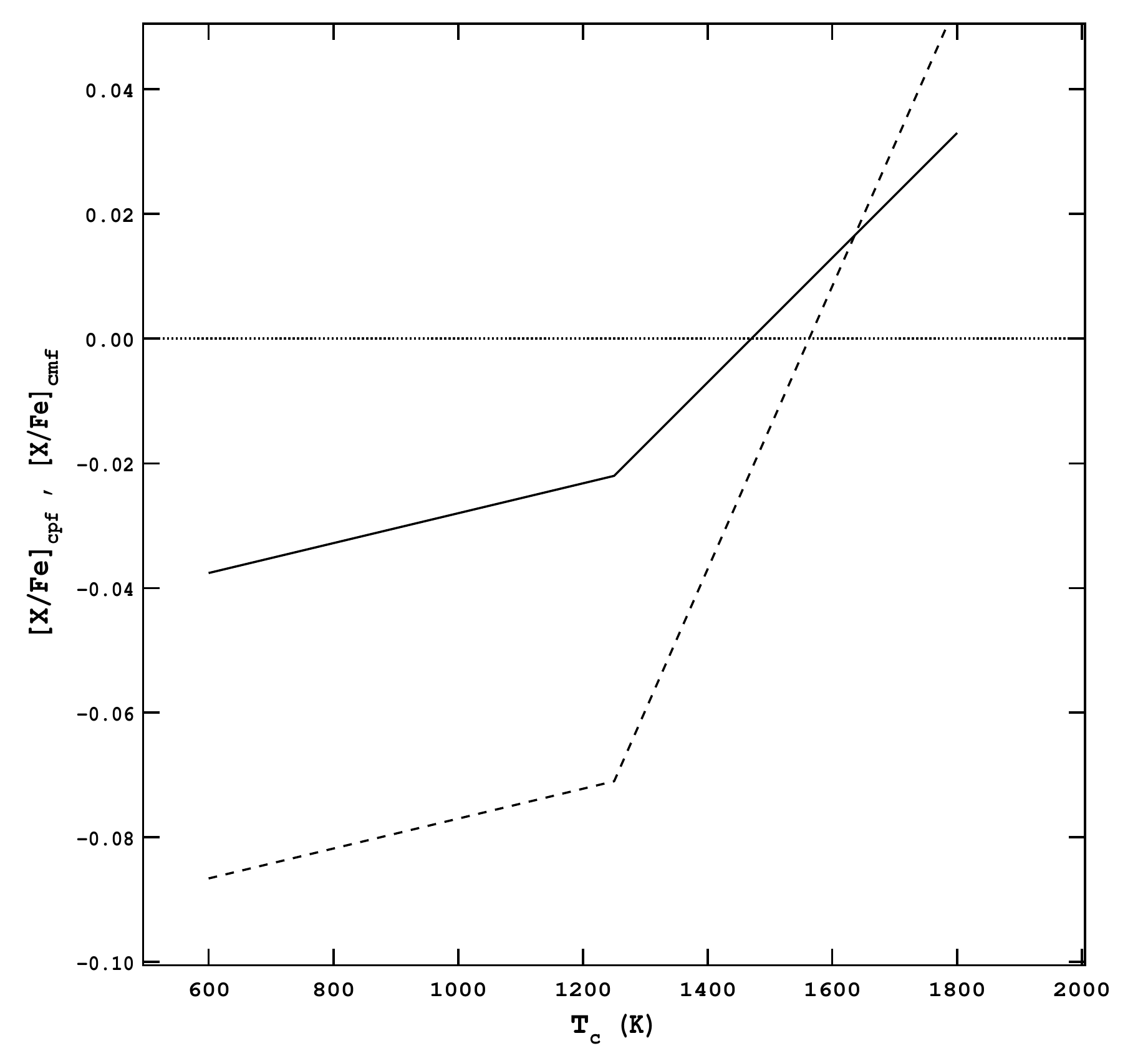}
 \caption{[X/Fe]$_{\rm cpf}$ (solid) and [X/Fe]$_{\rm cmf}$ (dashed) versus T$_{\rm c}$.}
\end{figure}

We show [X/Fe]$_{\rm cpf}$ and [X/Fe]$_{\rm cmf}$ plotted against T$_{\rm c}$ in Figure 3. In Table 1 we list the primordial meteoritic and photospheric abundances, calculated using the equations from this section. Only those elements included in the present analysis are listed.

\begin{table*}
\centering
\begin{minipage}{100mm}
\caption{Primordial Solar System photospheric and meteoritic abundances corrected for trends with T$_{\rm c}$. Abundances are expressed on a scale where $\log$ N(H) = 12.0. The weighted averages and weighted average sample errors are also listed; note, the weighted error value has been set to 0.01, even if the calculated weighted average error is smaller than 0.01.}
\label{xmm}
\begin{tabular}{lllll}
\hline
Element & Photosphere & Meteorites & Average & error\\
Na & 6.25 & 6.30 & 6.29 & 0.03\\
Mg & 7.65 & 7.59 & 7.59 & 0.04\\
Al & 6.52 & 6.56 & 6.55 & 0.03\\
Si & 7.57 & 7.57 & 7.57 & 0.01\\
P & 5.46 & 5.46 & 5.43 & 0.01\\
S & 7.15 & 7.17 & 7.17 & 0.01\\
K & 5.08 & 5.11 & 5.08 & 0.02\\
Ca & 6.41 & 6.42 & 6.42 & 0.01\\
Sc & 3.25 & 3.18 & 3.19 & 0.05\\
Ti & 4.98 & 5.02 & 5.01 & 0.03\\
V & 3.96 & 4.04 & 4.04 & 0.06\\
Cr & 5.67 & 5.69 & 5.69 & 0.01\\
Mn & 5.47 & 5.51 & 5.51 & 0.03\\
Fe & 7.53 & 7.50 & 7.51 & 0.02\\
Co & 4.99 & 4.93 & 4.93 & 0.04\\
Ni & 6.26 & 6.26 & 6.26 & 0.01\\
Cu & 4.22 & 4.28 & 4.26 & 0.04\\
Zn & 4.60 & 4.65 & 4.63 & 0.04\\
Ga & 3.06 & 3.11 & 3.10 & 0.03\\
Ge & 3.67 & 3.61 & 3.62 & 0.04\\
Rb & 2.51 & 2.38 & 2.40 & 0.09\\
Sr & 2.90 & 2.96 & 2.95 & 0.04\\
Y & 2.30 & 2.30 & 2.30 & 0.01\\
Zr & 2.69 & 2.68 & 2.69 & 0.01\\
Nb & 1.55 & 1.52 & 1.53 & 0.02\\
Mo & 1.96 & 2.05 & 2.04 & 0.06\\
Ru & 1.83 & 1.86 & 1.86 & 0.02\\
Rh & 0.95 & 1.13 & 1.09 & 0.12\\
Pd & 1.61 & 1.70 & 1.69 & 0.07\\
Ag & 1.00 & 1.23 & 1.22 & 0.16\\
Sn & 2.05 & 2.09 & 2.08 & 0.03\\
Ba & 2.32 & 2.26 & 2.27 & 0.04\\
La & 1.19 & 1.28 & 1.26 & 0.06\\
Ce & 1.65 & 1.67 & 1.67 & 0.01\\
Pr & 0.80 & 0.87 & 0.85 & 0.05\\
Nd & 1.50 & 1.56 & 1.55 & 0.04\\
Sm & 1.03 & 1.05 & 1.05 & 0.01\\
Eu & 0.58 & 0.57 & 0.57 & 0.01\\
Gd & 1.17 & 1.18 & 1.18 & 0.01\\
Tb & 0.40 & 0.45 & 0.44 & 0.03\\
Dy & 1.19 & 1.26 & 1.24 & 0.05\\
Er & 1.02 & 1.05 & 1.04 & 0.02\\
Tm & 0.20 & 0.25 & 0.23 & 0.03\\
Lu & 0.19 & 0.22 & 0.22 & 0.02\\
Hf & 0.94 & 0.85 & 0.86 & 0.07\\
Os & 1.50 & 1.51 & 1.51 & 0.01\\
Ir & 1.50 & 1.44 & 1.44 & 0.05\\
Au & 0.95 & 0.83 & 0.86 & 0.09\\
Pb & 1.96 & 2.06 & 2.05 & 0.08\\
Th & 0.12 & 0.19 & 0.18 & 0.05\\

\hline
\end{tabular}
\end{minipage}
\end{table*}

It is not surprising that the abundances of a number of elements remain discrepant between the photosphere and meteorites. The scatter evident in Figure 1 is mostly due to sources other than the systematic trend with $T_{\rm c}$. At least now we know the extent to which fractionation process have altered the Solar System meteoritic and photospheric abundances. A particularly notable change is the reduction in $\log$ (Pb/Sm) by 0.10 dex. This is more than enough to account for the excess in $^{204}$Pb relative to the model of \citet{bis14} in their recent $s$-process calculations. In any case, it will be necessary to perform a new fit to the Solar System $s$-process element distribution.

\section{Conclusions}

We have shown that the most recent published photospheric and meteoritic Solar System abundances display a significant trend with T$_{\rm c}$ when compared to each other, confirming previous findings. Both photospheric and meteoritic abundances display trends with T$_{\rm c}$ when solar analog stars abundances are employed as an external reference.

It has been assumed that CI chondrite meteorites are the least affected by fractionation processes, given their higher volatile/refractory ratio than other meteorite classes \citep{lod09}. However, it had not been demonstrated that the CI chondrites are completely lacking in fractionation effects for elements more refractory than H, C, N and O relative to the primordial state. If our finding is confirmed, then a mechanism for this fractionation will need to be sought.

\section*{Acknowledgments}

We thank the reviewer for very helpful comments and suggestions.

\bsp

\label{lastpage}


\begin{thebibliography}{}

\bibitem[\protect\citeauthoryear{Asplund et al.}{2005}]{asp05} Asplund M., Grevesse N., Sauval A. J., 2005, in Bash F. N., Barnes T. G., eds, ASP Conf. Ser., Cosmic Abundances as Records of Stellar Evolution and Nucleosynthesis, in press (astro-ph/0410214v2)
\bibitem[\protect\citeauthoryear{Asplund et al.}{2009}]{asp09} Asplund M., Grevesse N., Sauval A. J., Scott P., 2009, ARAA, 47, 481
\bibitem[\protect\citeauthoryear{Basu \& Antia}{2013}]{bas13} Basu S., Antia H. M., 2013, J Phys: Conf Ser, 440, id.012017
\bibitem[\protect\citeauthoryear{Bisterzo et al.}{2014}]{bis14} Bisterzo S., Travaglio C., Gallino R., Wiescher M., K\"appeler F., 2014, ApJ, 787, 10
\bibitem[\protect\citeauthoryear{Chambers}{2010}]{cham10} Chambers J., 2010, ApJ, 724, 92
\bibitem[\protect\citeauthoryear{Gonzalez}{1997}]{gg97} Gonzalez G., 1997, MNRAS, 285, 403
\bibitem[\protect\citeauthoryear{Gonzalez}{2006}]{gg06} Gonzalez G., 2006, MNRAS, 367, L37
\bibitem[\protect\citeauthoryear{Gonzalez et al.}{2010}]{gg10} Gonzalez G., Carlson M. K., Tobin R. W., 2010, MNRAS, 407, 314
\bibitem[\protect\citeauthoryear{Grevesse et al.}{2014}]{grev14} Grevesse N., Scott P., Asplund M., Sauval A. J., 2014, A\&A, in press.
\bibitem[\protect\citeauthoryear{Lodders}{2003}]{lod03} Lodders K., 2003, ApJ, 591, 1220
\bibitem[\protect\citeauthoryear{Lodders et al.}{2009}]{lod09} Lodders K., Palme H., Gail H. -P., 2009, Landolt-B\"ornstein -- Group VI Astronomy and Astrophysics Numerical Data and Functional Relationships in Science and Technology, Volume 4B: Solar System, J. E. Tr\"umper, ed., 4.4.
\bibitem[\protect\citeauthoryear{Melendez et al.}{2009}]{mel09} Melendez J., Asplund M., Gustafsson B., Yong D., 2009, ApJ, 704, L66
\bibitem[\protect\citeauthoryear{Melendez et al.}{2012}]{mel12} Melendez J., Bergemann M., Cohen J. G., Endl M., Karakas A. I., Ramirez I., Cochran W. D., Yong D., MacQueen P. J., Kobayashi C., Asplund M., 2012, A\&A, 543, A29
\bibitem[\protect\citeauthoryear{Mowlavi et al.}{2012}]{mow12} Mowlavi N., Eggenberger P., Meynet G., Ekstr\"om S., Georgy C., Maeder A., Charbonnel C., Eyer L., 2012, A\&A, 541, A41
\bibitem[\protect\citeauthoryear{Scott et al.}{2014a}]{scot14a} Scott P., Grevesse N., Asplund M., Sauval A. J., Lind K., Takeda Y., Collet R., Trampedach R., Hayek W., 2014, A\&A, in press.
\bibitem[\protect\citeauthoryear{Scott et al.}{2014b}]{scot14b} Scott P., Asplund M., Grevesse N., Bergemann M., Sauval A. J., 2014, A\&A, in press.



\end{thebibliography}
\end{document}